\documentclass[a4paper]{jpconf}
\usepackage{graphicx}
\usepackage{subfigure,graphicx,slashed,color}
\usepackage{wrapfig}
\usepackage{xspace,colortbl,times}
\usepackage{lmodern}
\usepackage{textcomp}
\usepackage[english]{babel}
\usepackage{tabularx}
\usepackage{amsmath}
\usepackage{amssymb}
\usepackage{fancyvrb}
\usepackage{fancyhdr}
\usepackage{bm}
\usepackage{framed}
\usepackage{booktabs}
\usepackage{alltt}
\usepackage{listings}
\usepackage{lscape}
\usepackage{rotating}
\usepackage{booktabs}
\usepackage{array}
\usepackage{rotating}
\usepackage{fancyhdr}
\usepackage{lineno}

\begin{document}

\pagestyle{plain}\pagenumbering{arabic}
\setcounter{page}{1}

\pagestyle{fancy}
\fancyhf{}
\cfoot{ \thepage}


\newcommand{\pip}{\ensuremath{\pi^{+}}}
\newcommand{\pim}{\ensuremath{\pi^{-}}}
\newcommand{\piz}{\ensuremath{\pi^{0}}}
\newcommand{\kap}{\ensuremath{{\rm K}^{+}}}
\newcommand{\kam}{\ensuremath{{\rm K}^{-}}}
\newcommand{\pbar}{\ensuremath{\rm\overline{p}}}
\newcommand{\sqrtS}{\ensuremath{\sqrt{s}}}
\newcommand{\pp}{\ensuremath{\mathrm {p\kern-0.05emp}}}
\newcommand{\PbPb}{\ensuremath{\mbox{Pb--Pb}}}
\newcommand{\pPb}{\ensuremath{\mbox{p--Pb}}}
\newcommand{\dEdx}{\ensuremath{\mathrm{d}E/\mathrm{d}x}}
\newcommand{\dndy}{\ensuremath{\mathrm{d}N/\mathrm{d}y}}

\title{Production measurements of heavy quarks in pp collisions
	at  $\sqrt{s}$ = 13 TeV  with the ALICE detector}

\author{Tebogo Joyce Shaba, for the ALICE Collaboration\\ North-West University, South Africa \\ iThemba LABS, Sommerset West, Western Cape }

\ead{tebogojoyce.shaba@cern.ch}

\begin{abstract}Heavy-flavour production measurements in pp collisions are
important tools to test theoretical models based on perturbative
quantum chromodymanics (pQCD) and to investigate the heavy-quark hadronization mechanisms. In ALICE, heavy quarks are
measured via the hadronic and electronic decay channels at central
rapidity (-0.9 $<$ $\textit{y}$ $<$ 0.9) and via the muon decay channels at forward
rapidity (-4 $<$  $\textit{y}$ $<$ -2.5).\newline 

In this contribution, the production cross-section measurements via the leptonic decay of heavy-flavour hadrons are presented and
compared to pQCD theoretical
calculations. The latest measurements of  D$\textsuperscript{0}$,  D$\textsuperscript{+}$,  D$\textsuperscript{*+}$,   $D^{+}_{s}$ mesons whose hadronic decays into charged are fully reconstructed
together with the measurements of $\Lambda^{+}_{c}$ , $\Xi^{0,+}_{c}$, $\Sigma^{0,++}_{c}$ and $\Omega^{0}_{c}$ baryons, performed with the ALICE detector at
midrapidity in pp collisions at $\sqrt{s}$ = 13 TeV, are also presented.
Measurements of charm-baryon production are crucial to study the
charm-quark hadronization mechanisms in a partonic rich environment
like the one produced in pp collisions at LHC energies.
\end{abstract}

\section{Introduction}
Heavy-flavours (charm and beauty) are abundantly produced at the LHC \cite{LHC} in the early stages of hadronic collisions via hard parton-parton scattering processes and therefore experience the full evolution of the system. The measurement of heavy flavours in \pp\ collisions can be used to test pQCD calculations. In addition, these measurements provide an  essential baseline for the studies of nuclear (\pPb\ and \PbPb) collisions. Heavy flavour production in nuclear collisions is modified by cold nuclear matter effects (CNM) such as shadowing and energy loss \cite{eloss, cnm}. The knowledge of these effects is fundamental for understanding the interactions of heavy quarks with the deconfined medium formed in heavy-ion collisions where the modified transverse momentum (\textit{p}$_T$) distribution can be used to infer the effect of the interactions with the QGP. Heavy-quark production is experimentally accessible through the measurement of heavy-flavour hadrons and their decay products. \newline

At the LHC, ALICE \cite{alice} comprises several subdetectors, which are used for the measurement of charged hadrons, charged leptons and photons, including forward muon spectrometer which measures muons. The L3 solenoid magnet provides a field of 0.5 T to the central barrel detectors covering the rapidity -0.9 $<$ $\textit{y}$ $<$ 0.9. The Silicon Pixel Detector (SPD), as part of the inner tracking system, is used for the determination of the interaction vertex as well as the measurement of charged-particle multiplicity. Particle-identification information is given by the Time Of Flight (TOF) and the Time Projection Chamber (TPC).   Two arrays of scintillator detectors (V0A and V0C),  placed on both sides of the interaction point  provide centrality in \PbPb\ collisions as well as trigger information and beam-gas background suppression. They are also used for multiplicity determination. The Transition Radiation Detector (TRD), is used for the electron identification through the ionization energy and the Electromagnetic Calorimeter (EMCal) that is the last detector before the magnet is used to improve ALICE capacity in \textit{p}$_T$ reconstruction of jets, direct photons and electrons from heavy flavor decay. The forward muon spectrometer covers the rapidity range -4 $<$  $\textit{y}$ $<$ -2.5. It consists of a composite absorber, a dipole magnet, five tracking stations, a muon filter and two trigger stations. The absorber reduces background muons mainly from the decays of pions and kaons. The dipole magnet  provides a horizontal magnetic field perpendicular to the beam axis and is used for charge and momentum determination. The tracking stations are used to determine the trajectories of the muons traversing the detector. The muon filter is mounted in front of the trigger stations and filters out all particles with momentum $\textit{p}$ $<$ 4 GeV/c. The trigger stations are used for muon identification and triggering.\newline
 
The reconstruction of charm  baryons and mesons is done via their hadronic decay channels. Table \ref{tab1} below shows the decay channels and their branching ratios (BR)\newline

\begin{table}[!ht]
	\centering
	
	\begin{tabular}{ |c|c| } 
		\hline
$	\textbf{	Charm mesons} $& $\textbf{Charm Baryons}$  \\ 
		\hline 
		D$\textsuperscript{0}$ $\rightarrow$ \kam\pip (3.88$\pm$0.05 \%) & $\Lambda^{+}_{c}$ $\rightarrow$ $K^{+}_{c}$p (1. 59$\pm$  0.08\%)  \\ 
		D$\textsuperscript{+}$ $\rightarrow$ \kam\pip\pip (9.13$\pm$0.19 \%) & $\Xi^{0}_{c}$ $\rightarrow$ $\Xi^{-}$$\pi^{-}$ (1.43$\pm$ 0.32 \%)  \\ 
	$D^{*+}$ $\rightarrow$ $D^{0}$\pip\ (67.7$\pm$0.05 \%) & $\Sigma^{0,++}_{c}$ $\rightarrow$ $\Lambda^{+}_{c}$ $\pi^{-,+}$ ( $\approx$ 100\%)\\
	$D^{+}_{s}$ $\rightarrow$ \kam\kap\pip ( $\approx$ 5.39\%) & $\Omega^{0}_{c}$ $\rightarrow$ $\Omega^{-}$$\pi^{+}$ (0.51$\pm$2.19 \%)\\
		 & \\
		\hline
	\end{tabular}\newline

\caption{The hadronic decay modes of charm mesons and charm baryons exploited by ALICE with in parenthesis the respective BR. \label{tab1}}.

\end{table}

 \
 \section{Results}

\subsection{$\textbf{Charm baryon production cross sections}$}
\begin{figure}[!h]
	\centering
		\subfigure[]
		{
				\includegraphics[width=200px,height=190px]{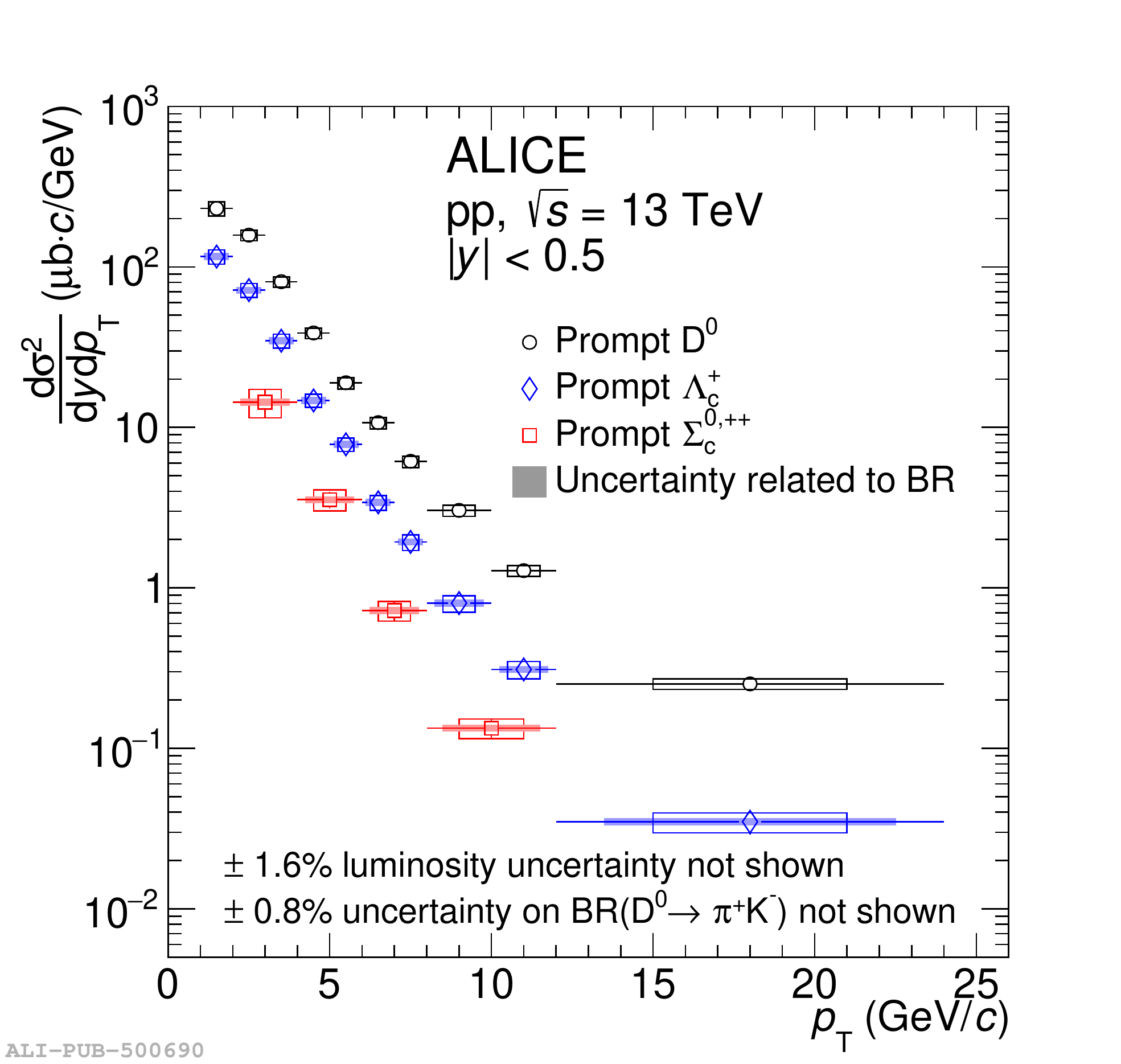}
				\label{fig:first_4_1}
			}
			\subfigure[]
		{
			\includegraphics[width=230px,height=180px]{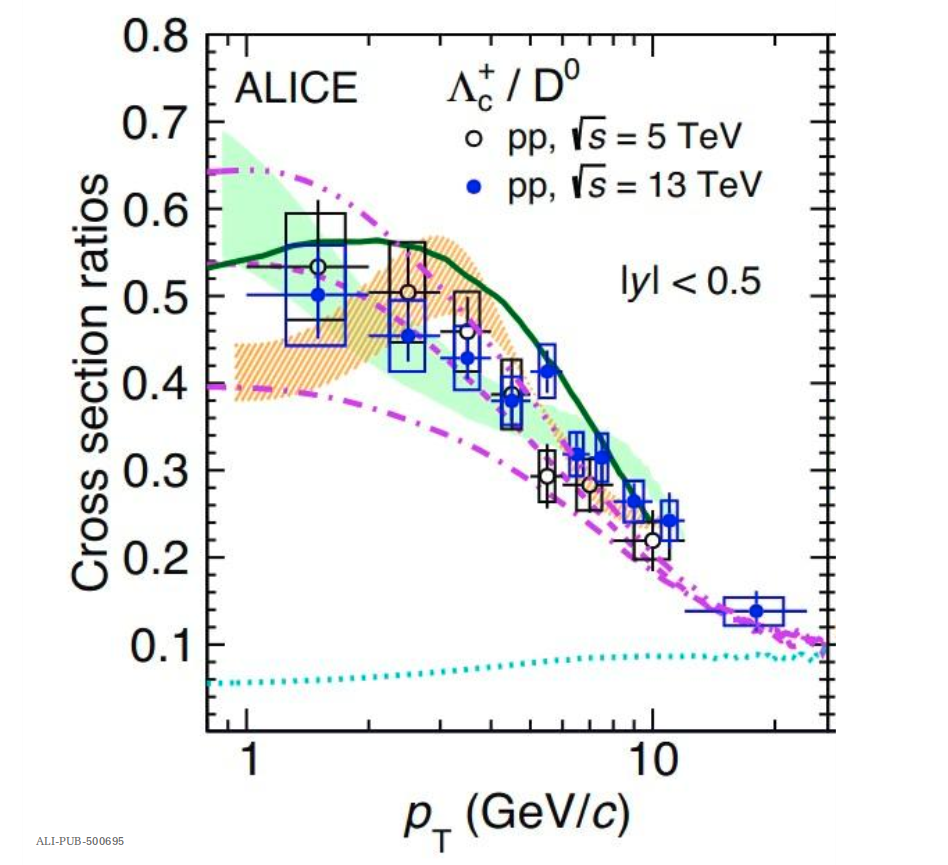}
			\label{fig:first_4_2}
		}	
		
		\caption{\textit{p}$_T$-differential cross sections of prompt D$\textsuperscript{0}$, $\Lambda_{c}$, and $\Sigma^{0,++}_{c}$ (left)  and prompt-charm-hadron cross-section ratios (right) at $\sqrt{s}$ = 13 TeV  compared with data from pp collisions at \pp\ collisions at $\sqrt{s}$ = 5 TeV  \cite{FIG2a}.}
		\label{results3}
\end{figure}

Figure \ref{results3}a and 1b  \cite{FIG2a}  respectively, shows the \textit{p}$_T$-differential cross section of prompt D$\textsuperscript{0}$ and charm baryons: $\Lambda^{+}_{c}$, $\Sigma^{0,++}_{c}$ (left), and prompt-charm-hadron cross-section ratio of $\Lambda_{c}$/D$\textsuperscript{0}$ (right) in pp collisions at $\sqrt{s}$ = 13 TeV. The baryon-to-meson yield ratios are taken into consideration to probe different hadronisation mechanisms. The baryon-to-meson yield ratios are higher in pp collisions at LHC energies than in e$\textsuperscript{+}$e$\textsuperscript{-}$ collisions, suggesting that charm hadronisation occurs via different processes in the two collision systems. The $\Lambda_{c}$/D$\textsuperscript{0}$ ratio decreases with increasing \textit{p}$_T$ and has a strong enhancement for \textit{p}$_T$ $<$  10  GeV/c. The values measured in pp collisions at $\sqrt{s}$ = 13 TeV are compatible, within uncertainties, with those measured at $\sqrt{s}$ = 5 TeV.

\subsection{$\textbf{Heavy charm baryons $\Sigma^{0,++}_{c}$, $\Xi^{+}_{c}$ }$}
In Figure 2 , the  $\Sigma^{0,++}_{c}$/D$\textsuperscript{0}$ (left)  \cite{FIG2a} ratios shows a remarkable difference between the pp and e$\textsuperscript{+}$e$\textsuperscript{-}$ collisions by a factor of approximately 10. In Figure 2, the data are descibed well by the models, except for PYTHIA8 Monash Tune \cite{PYTHIA8monash}. The baryon-to-meson ratio (right) \cite{FIG3b} increases towards low \textit{p}$_T$ up to a value of approximately 0.3, while PYTHIA8 Monash, PYTHIA8 CR Tune \cite{PYTHIA8CR}, SHM+RQM \cite{SHMRQM} and QCM \cite{QCM} significantly underestimate the ratios. The Catania  model \cite{CATANIA} describes better the ratios in the mearsured  \textit{p}$_T$ interval. 
\begin{figure}[!h]
	\begin{center}
		
			\subfigure[]
		{
		\includegraphics[width=200px,height=180px]{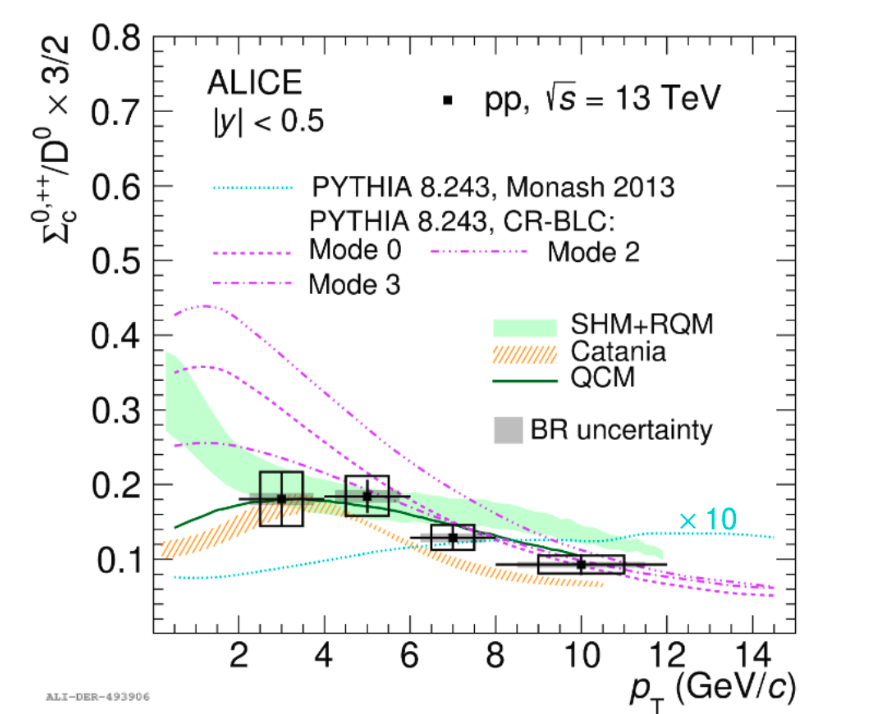}
	}
			\subfigure[]
		{
			\includegraphics[width=220px,height=190px]{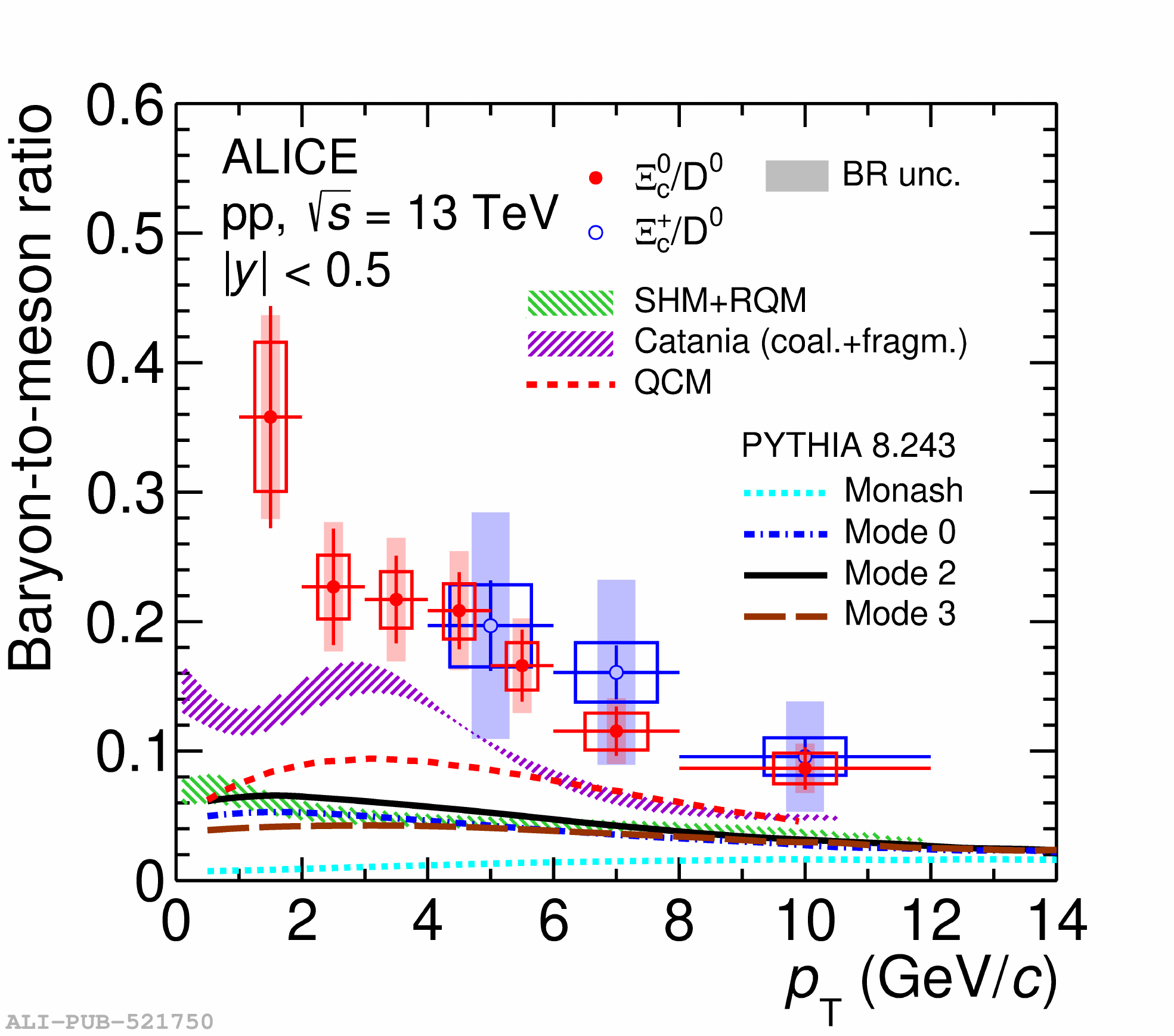}
			\label{fig:first_4_2}
		}

				\caption{The prompt-charm-hadron cross section ratios (right) for  $\Sigma^{0,++}_{c}$/D$\textsuperscript{0}$ (left)  \cite{FIG2a} and $\Xi^{0}_{c}$/D$\textsuperscript{0}$ and $\Xi^{+}_{c}$/D$\textsuperscript{0}$ ratio (right) \cite{FIG3b} as a function of  \textit{p}$_T$ in \pp\ collisions at $\sqrt{s}$ = 13 TeV }
		\label{results5}
	\end{center}
\end{figure}

	
	

\subsection{$\textbf{Heavy charm baryons $\Omega^{0}_{c}$}$}
In Figure 3 \cite{FIG4}, the measurement of $\Omega^{0}_{c}$ is compared with model predictions of the PYTHIA 8 Monash tune and with CR (colour reconnections) beyond the leading-colour approximation \cite{PYTHIA8CR}, which are multiplied by a theoretical BR ($\Omega^{0}_{c}$ $\rightarrow$ $\Omega^{+}$$\pi^{-}$). The cross section from the CR-BLC tune \cite{CRBLC} is larger than the one from the Monash tune by a factor varying between 9 and 25 depending on \textit{p}$_T$.

\begin{figure}[!h]
	\centering
	
	{
		\includegraphics[width=200px,height=180px]{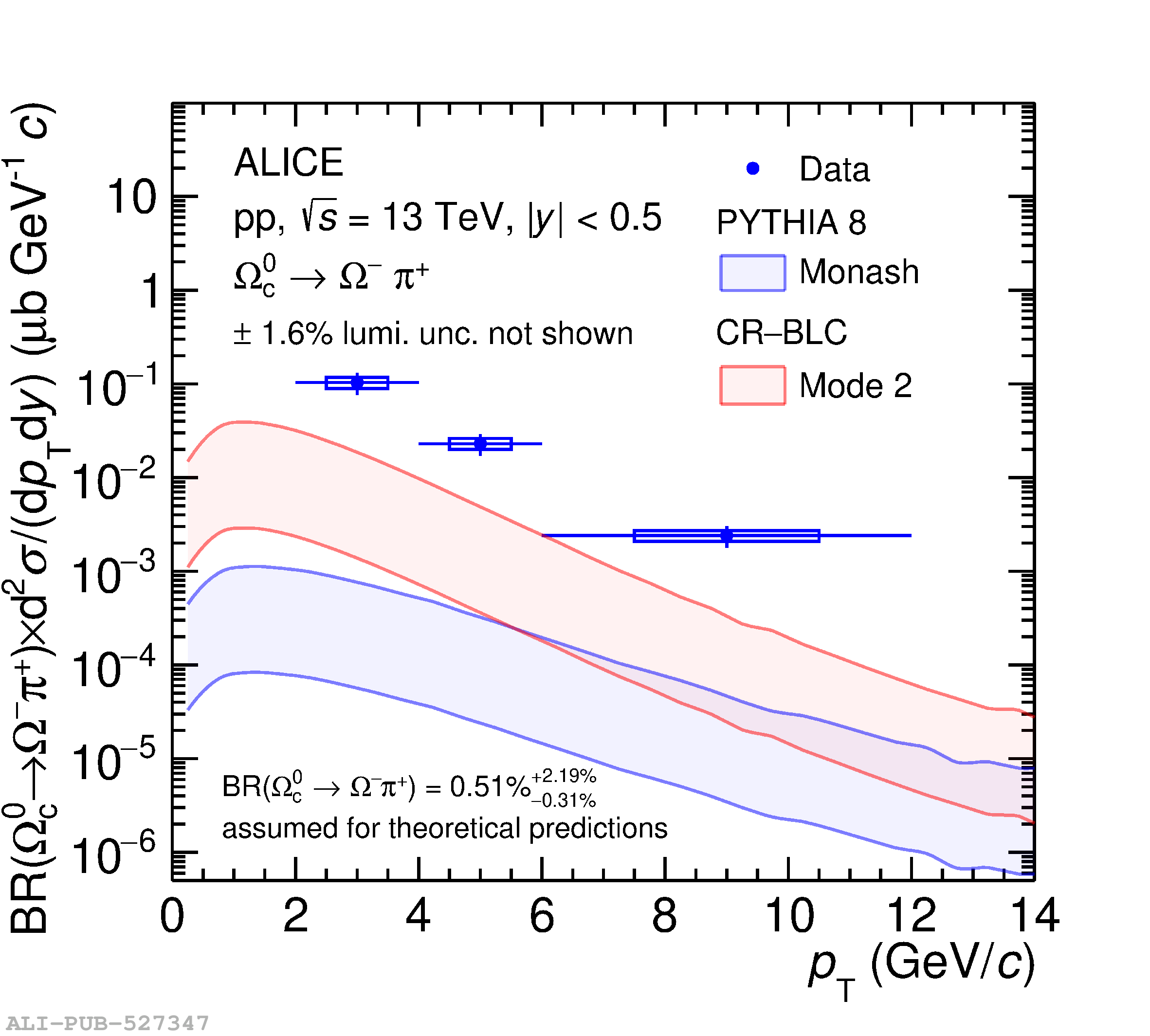}
		\label{fig:first_4_1}
	}
	
	\caption{The \textit{p}$_T$-differential production cross section of inclusive $\Omega^{0}_{c}$ baryons multiplied by the branching ratio into $\Omega^{+}$$\pi^{-}$   in \pp\ collisions at $\sqrt{s}$ = 13 TeV \cite{FIG4}.}
	\label{results4_1}
	
\end{figure}

		


\section{Conclusion}
 PYTHIA 8 Monash, which adopts fragmentation functions constrained in $\e^{+}$$\e^{-}$ and $\e^{+}$p collisions, significantly underestimate the baryon-to-meson productions, while models with an
augmented formation of baryons (PYTHIA 8 CR tunes, SHM+RQM, Catania) are closer to the data and can describe $\Lambda_{c}$ and $\Sigma_{c}$, for the strange charm baryons ($\Omega_{c}$, $\Xi_{c}$). All the models underestimate the data, with the Catania model providing the closest agreement. The non-prompt $\Lambda^{+}_{c}$/D$\textsuperscript{0}$ provides indirect access to the beauty fragmentation. The \textit{p}$_T$ trend of prompt and non-prompt $\Lambda^{+}_{c}$/D$\textsuperscript{0}$ is similar.

\end{document}